\documentclass{article}
\usepackage{spconf,amsmath,graphicx,hyperref,booktabs, array, etoolbox}
\usepackage{xcolor}
\usepackage{amsmath,amssymb,amsfonts}
\usepackage{spconf,amsmath,graphicx,hyperref,orcidlink}
\usepackage{amsmath,amssymb,amsfonts}
\usepackage{algorithmic}
\usepackage{graphicx}
\usepackage{textcomp}
\usepackage{xcolor}
\usepackage{url}
\usepackage{comment}
\usepackage{tabularx}
\usepackage{booktabs}
\usepackage{array}
\usepackage{multirow}
\usepackage{tikz}
\usepackage{amssymb}
\usepackage[numbers]{natbib}
\usepackage{setspace}

\title{AFA-Net: A Differential Attention Approach for Auditory Attention Detection}

\name{Philip H. Lee\orcidlink{0009-0006-9090-3394}$^{*}$, Shreeram Suresh Chandra\orcidlink{0000-0002-1085-2544}$^{\dag}$, Karan Thakkar\orcidlink{0000-0002-6803-1116}$^{\ddag}$, John H.L. Hansen\orcidlink{0000-0003-1382-9929}$^{\S}$}

\address{$^{*}$Independent Researcher, USA \\
         $^{\dag}$Center for Language and Speech Processing (CLSP), Johns Hopkins University, USA \\
         $^{\ddag}$Laboratory for Computational Audio Perception (LCAP), Johns Hopkins University, USA \\
         $^{\S}$Center for Robust Speech Systems (CRSS), University of Texas at Dallas, USA}
         
\begin{document}
%
\maketitle
\begin{abstract}
    Auditory Attention Detection (AAD) utilizes electroencephalographic (EEG) signals to identify a target speaker in a multi-speaker environment. Despite considerable progress, existing deep learning architectures often lack explicit mechanisms for handling noisy EEG data. To address this limitation, we propose Auditory Focus Attention Networks (AFA-Net), a machine learning framework that replaces vanilla attention with a simple yet flexible differential attention mechanism to help focus on task-relevant neural activity. AFA-Net achieves an upward accuracy of 96.8\% at the 2s decision window, while using substantially fewer parameters than most existing methods. To the best of our knowledge, AFA-Net is among the first frameworks to explicitly try to combat EEG noise to improve AAD.
\end{abstract}
\vspace{-10pt}
\section{Introduction}
The \textit{cocktail party effect} refers to the ability to focus on a specific voice while filtering out background noise \cite{wen2025hybrid}. This capability is crucial for people with hearing impairments, as they often struggle to distinguish target speakers in noisy environments. Given the established link between auditory attention and brain activity, prior work has explored several neural modalities for AAD, including EEG \cite{OSullivan2015, Mirkovic2015}, ECoG \cite{Mesgarani2012, Golumbic2013}, and MEG \cite{Ding2012, Akram2016}. Given EEG's cost-effectiveness and non-invasive approach, it has emerged as the popular choice for AAD, albeit its issues with containing channel and recording noise.

Earlier methods used Convolutional Neural Networks (CNNs) to capture local patterns \cite{vandecappelle2021eeg, Li2025MHANet} in AAD. Transformers were then later favored for capturing global dependencies through self-attention \cite{Su2022, Kuruvila2021}, and hybrid CNN-transformer architectures have since achieved state-of-the-art (SOTA) performance by combining both strengths \cite{Yan2024DARNet, Zhu2023}. While these advances have improved AAD accuracy across short decision windows, existing methods do not attempt to explicitly address the noise found in EEG signals.

A key limitation in current approaches stems primarily from the design of the attention mechanism. Vanilla self-attention utilizes softmax normalization, forcing positive attention weights to be distributed across all EEG features \cite{ICLR2025_00b67df2}. As a result, self-attention at best can only dilute rather than eliminate irrelevant or spurious details. This is especially problematic for EEG signals, which are prone to artifacts and background activity \cite{Li2024Attention, Li2024RL}. In response to this limitation, we adopt a differential attention mechansim \cite{ICLR2025_00b67df2}, which creates and subtracts two attention maps, giving it the potential to assign zero or even negative attention weights to unimportant features. In doing so, the overall attention mechanism is freed from the constraint of having to account for every feature, allowing it to focus more deliberately on neural activity tied to the attended speaker.

\begin{figure*}[t]
    \centering
    \includegraphics[width=\textwidth]{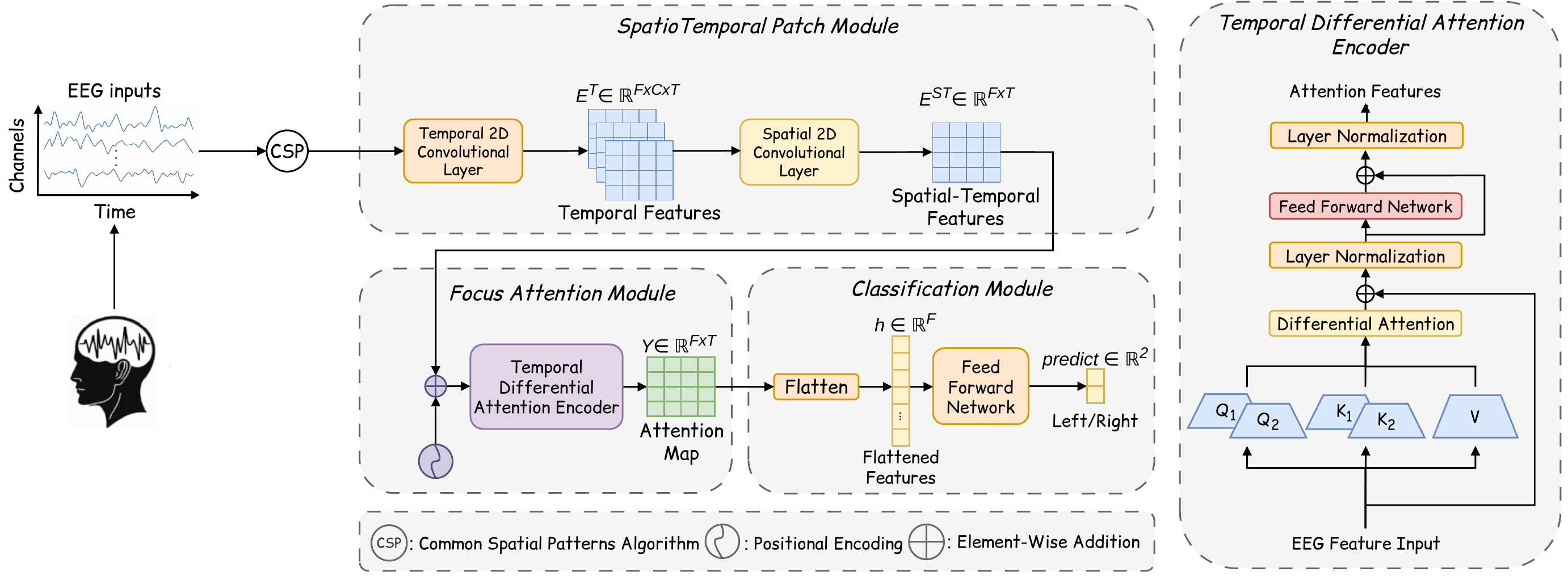}
    \caption{Proposed architecture of AFA-Net.}
    \label{fig:framework}
   
\end{figure*}

Building on this idea, we propose AFA-Net, a lightweight and novel architecture that leverages differential attention \cite{ICLR2025_00b67df2} to achieve more accurate AAD. Our model consists of two modules: 1) the \textit{SpatioTemporal Patch Module} (STPM), which uses 2D CNNs to extract local spatio-temporal features, and 2) the \textit{Focus Attention Module} (FAM), which incorporates differential attention \cite{ICLR2025_00b67df2} to capture relevant global EEG features. 

The overall contributions are as follows: 1) we propose AFA-Net, a novel architecture that helps distinguish relevant neural signals from unimportant information for AAD; 2) we show that differential attention improves accuracy while adding only a negligible increase in parameter count; 3) AFA-Net achieves competitive performance across all decision windows; and 4) we conduct an architectural ablation validating each module's contribution and a frequency band ablation showing that differential attention attends more effectively to the frequencies most informative for AAD compared to vanilla attention.

\vspace{-10pt}
\section{Main Method}
We propose AFA-Net, a differential attention-based architecture that emphasizes relevant EEG features for improved AAD performance. As shown in Figure \ref{fig:framework}, AFA-Net incorporates two modules: STPM, which extracts local EEG embeddings, and FAM, which extracts global EEG features.

Before inputting EEG data into AFA-Net, raw EEG signals are first segmented into fixed-length time decision windows. We then preprocess each window using the CSP algorithm \cite{ramoser2000optimal, blankertz2007optimizing} to maximize class separation between attended and unattended states. Let $E^{CSP} \in \mathbb{R}^{C \times T}$ represent the EEG segment within a single decision window, where $C$ is the number of channels and $T$ is the number of timestamps in each window.

\vspace{-8pt}
\subsection{SpatioTemporal Patch Module}
EEG signals contain meaningful spatial and temporal structure for AAD. Motivated by their effectiveness in capturing local patterns \cite{Yan2024DARNet}, we use 2D convolutional layers to extract these features, beginning with two temporal convolutional layers to capture short-range relationships across overlapping timestamps. This is represented as 
\begin{equation}
\begin{aligned}
E^{T}_{1} &= \mathrm{GELU}(\mathrm{TemporalConv2D}(E^{CSP})) \\
E^{T}_{2} &= \mathrm{GELU}(\mathrm{TemporalConv2D}(E^{T}_{1})),
\end{aligned}
\end{equation} where $E^{T}_{1}, E^{T}_{2} \in \mathbb{R}^{F \times C \times T}$ are the outputs of the first and second temporal convolutional layer, with $F$ denoting the embedding dimension. $TemporalConv2D(.)$ applies a 2D convolutional filter with a kernel size (1, $k_\text{time}$), where $k_\text{time}$ is the total timestamps, $T$, along the time axis. A GELU activation follows each convolutional layer to introduce nonlinearity.

Following temporal feature extraction, we incorporate a single spatial convolutional layer to capture cross-channel relationships. This is formulated below as
\begin{equation}
\begin{aligned}
E^{ST} &= \mathrm{GELU}(\mathrm{SpatialConv2D}(E^T_2)),
\end{aligned}
\end{equation} where $E^{ST} \in \mathbb{R}^{F \times T}$ represents the spatial-temporal feature output of STPM. Here, $SpatialConv2D(.)$ applies a 2D convolutional filter with kernel size ($k_\text{space}$, 1), where $k_\text{space}$ is set to the total number of electrode channels, $C$, present. This helps aggregates channel-related information, resulting in the final spatial-temporal before entering FAM.

\subsection{Focus Attention Module}
Before applying differential attention, we first add absolute positional embeddings \cite{vaswani2017attention} to $E^{ST}$ to preserve temporal ordering, to yield the output $\hat{X} \in \mathbb{R}^{T \times F}$. We then apply differential attention as shown in Figure \ref{fig:framework}. We begin by linearly projecting $\hat{X}$ onto query, key, and value vectors. The query and key embeddings are split so that their feature dimension $F$ is cut in half, yielding $Q_1, Q_2, K_1, K_2 \in \mathbb{R}^{T \times \frac{F}{2}}$, while $V \in \mathbb{R}^{T \times F}$ remains the same size.  These projections are computed as follows:
\begin{equation}
[Q_1; Q_2] = \hat{X}W^Q, \enspace [K_1; K_2] = \hat{X}W^K, \enspace V = \hat{X}W^V,
\end{equation}
where $W^{Q}, W^{K}, W^{V} \in \mathbb{R}^{F \times F}$ are the corresponding weight matrices. Cutting the feature dimension in half in this way allows two separate attention maps to be computed independently with negligible computational increases. We formulate the attention calculations below, one from each query-key pair, ($Q_1, K_1$) and ($Q_2, K_2$):
\begin{equation}
A_1 = \mathrm{softmax}\left(\frac{Q_1 K_1^\top}{\sqrt{d/2}}\right), A_2 = \mathrm{softmax}\left(\frac{Q_2 K_2^\top}{\sqrt{d/2}}\right).
\end{equation} We then subtract $A_2$ from $A_1$ to obtain the final differential attention map, where $\lambda$ controls the degree to which the second attention map is subtracted. The $\lambda$ and differential attention are represented as:
\begin{equation}
\begin{gathered}
\lambda = \exp(\lambda_{q_1} \cdot \lambda_{k_1}) - \exp(\lambda_{q_2} \cdot \lambda_{k_2}) + \lambda_{\text{init}} \\[6pt]
\mathrm{DiffAtt}(\hat{X}) = (A_1 - \lambda A_2)V,
\end{gathered}
\end{equation} where $\lambda_{q_1}, \lambda_{k_1}, \lambda_{q_2}, \lambda_{k_2} \in \mathbb{R}^{F}$ are learnable weights with $\lambda_{\text{init}} \in (0,1)$ being a hyperparameter constant used for initialization. The resulting differential attention map can potentially span a wider range of attention weight values, which can potentially be negative.

We further use multiple heads \cite{vaswani2017attention} to capture attention patterns across the EEG data, then concatenate and project the outputs as follows:

\begin{equation}
\mathrm{MHA}(\hat{X}) = \left[\mathrm{DiffAtt}_1(\hat{X}), \cdots, \mathrm{DiffAtt}_H(\hat{X})\right]W^O,
\end{equation} where each head independently computes differential attention, jointly attending to EEG feature across channels and time. The output is then refined using a feed-forward network (FFN). Following the exact design of the FFN sublayer in vanilla transformers \cite{vaswani2017attention}, we apply two linear layers with ReLU activation alongside layer normalization and residual connections:
\begin{equation}
\begin{gathered}
Z = \mathrm{LayerNorm}(\hat{X} + \mathrm{MHA}(\hat{X})) \\[6pt]
\hat{Z} = \mathrm{LayerNorm}(Z + \mathrm{FFN}(Z)).
\end{gathered}
\end{equation}
The residual connections and $\mathrm{LayerNorm}(\cdot)$ stabilize training, yielding the final output $\hat{Z}$.

\subsection{Classification Layer}
The final classification layer maps $\hat{Z}$ to an attended speaker. After flattening, $Y$ is passed through a hidden layer with batch normalization, followed by a linear classifier:
\begin{equation}
\begin{gathered}
Y = Flatten(\hat{Z}) \\[6pt]
h = \mathrm{BN}(\mathrm{ReLU}(W_3 Y + b_3)) \\[6pt]
predict = W_4 h + b_4.
\end{gathered}
\end{equation}
Here, $W_3, W_4$ are learnable weight matrices and $\mathrm{BN(.)}$ and $\mathrm{ReLU(.)}$ denote batch normalization and ReLU activation. The output $predict \in \mathbb{R}^{2}$ gives the binary decision of the attended speaker.

\vspace{-8pt}
\section{Experimental Setup}
\vspace{-4pt}
\subsection{Datasets}
Experiments were conducted on widely-used baseline datasets: KUL \cite{das2019auditory} and DTU \cite{fuglsang2018eeg}.

In KUL \cite{das2019auditory}, 16 normal-hearing participants were recorded with a 64-channel BioSemi ActiveTwo system. They attended to one of two competing Dutch stories under dichotic presentation and HRTF-filtered spatial conditions simulating speech at $90^{\circ}$ to the left or right of the listener (8 trials, 6 min each).

In DTU \cite{fuglsang2018eeg}, 18 normal-hearing participants were recorded with a 64-channel BioSemi ActiveTwo system. They attended to a target speaker positioned at $60^{\circ}$ azimuth against a competing speaker, listening to Danish audiobooks (60 trials, 50 s each).

\vspace{-10pt}
\subsection{Data Preprocessing}
To ensure a fair comparison, similar preprocessing is applied to both datasets. For KUL, signals are re-referenced to the average mastoid, bandpass filtered (0.1–50 Hz), and downsampled to 128 Hz. For DTU, line noise and harmonics are removed, ocular artifacts are eliminated via joint decorrelation, and signals are downsampled to 128 Hz.

\vspace{-10pt}
\subsection{Training Configurations}
We evaluate AFA-Net on the KUL and DTU datasets using classification accuracy as the primary metric. Results for prior methods are obtained by retraining each model using its publicly available codebase, following its original preprocessing and training protocol \cite{cai2021low, Su2022, xu2024densenet}. For models employing CSP-based feature extraction \cite{ni2024dbpnet, Yan2024DARNet, Li2025MHANet}, we adopt the corrected experimental setup of \cite{Yan2024DARNet}. First, each trial is split temporally with first 90\% for training and the last 10\% for testing. Secondly, CSP feature extraction is only fit on the training data and then transformed on both splits. Finally, a sliding window with 50\% overlap is applied separately to the training and test portions, with training features further split into training (90\%) and validation (10\%) sets. AFA-Net is trained with AdamW optimizer with a learning rate of $1\times10^{-5}$ and a weight decay of $3\times10^{-4}$. During training, we also utilize a batch size of 32 for up to 100 epochs. We also apply early stopping after 10 epochs without validation improvement. To determine the significance of our results, we aggregate trial-level predictions into per-subject test accuracy. We then assess a subject-wise two-tailed paired t-test against the strongest baseline, MHANet \cite{Li2025MHANet}, with results and significant differences (marked as *) shown in Table \ref{tab:results}.

\begin{figure}[t]
  \centering
  \includegraphics[width=1.0\linewidth]{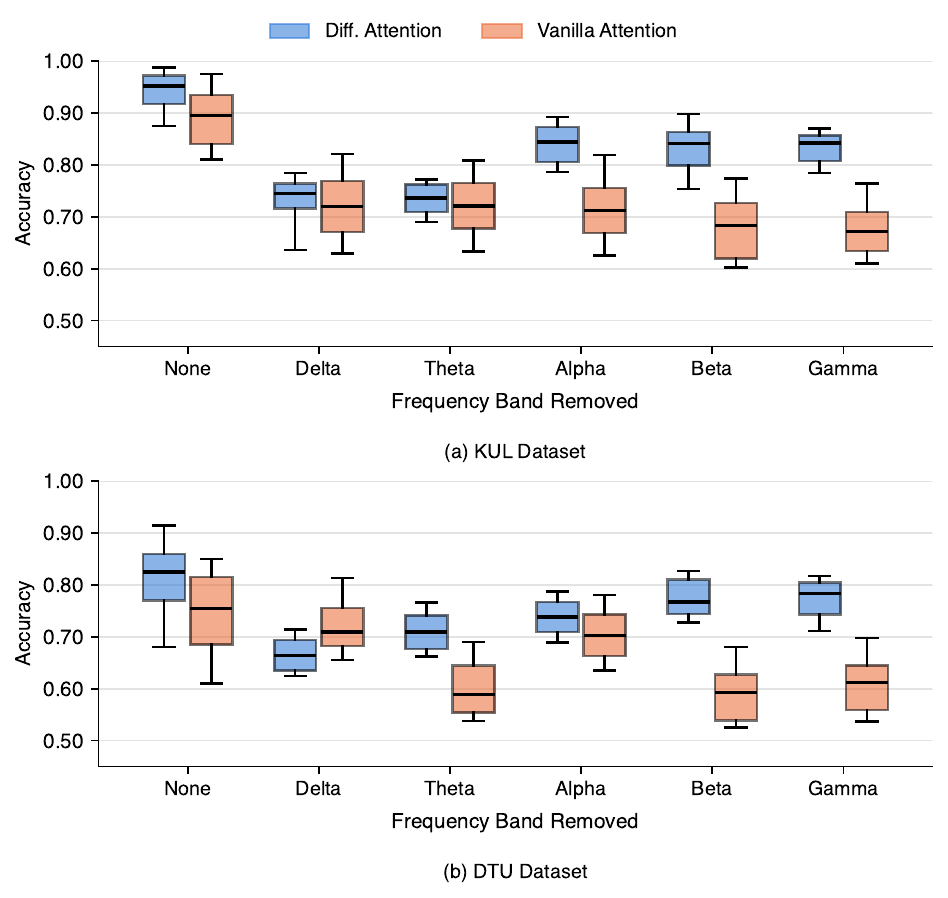}
  \caption{Ablation study comparing frequency band removal for differential vs. vanilla attention at the 1s decision window.}
  \label{fig:ablation-bands}
\end{figure}

\vspace{-12pt}
\section{Results}
\subsection{Performance Analysis}
We evaluate AFA-Net across 0.1s, 1s, and 2s decision windows on KUL and DTU, comparing against SOTA AAD models retrained under our corrected experimental setup (Table~\ref{tab:results}).

AFA-Net with differential attention achieves strong performance, as high as 96.8\% on KUL at the 2s decision window, using only 0.03M parameters. This is fewer than most baselines and comparable to DARNet (0.08M) and MHANet (0.02M). Despite the extra computation, differential attention adds negligible parameter count over vanilla attention (0.03M), while delivering higher accuracy. It also significantly outperforms MHANet, the strongest baseline, across all windows except at the 1s decision window on DTU.

\begin{table*}[t]
\centering
\caption{Comparison of auditory attention detection accuracy (\%) across different models on the KUL and DTU datasets using various decision windows. The * symbol denotes significant differences over the strongest baseline, MHANet (two-tailed paired t-test, $p < 0.05$).}
\label{tab:results}
\setlength{\tabcolsep}{2.2pt}
\renewcommand{\arraystretch}{0.7}
\begin{tabular}{l>{\centering\arraybackslash}p{1.4cm}cccccc}
\toprule
& & \multicolumn{3}{c}{\textbf{KUL}} & \multicolumn{3}{c}{\textbf{DTU}} \\
\cmidrule(lr){3-5} \cmidrule(lr){6-8}
Model & \text{Params (M)} & 0.1s & 1s & 2s & 0.1s & 1s & 2s \\
\midrule
SSF-CNN \cite{cai2021low}      & 4.21 & $76.3 \pm 8.47$ & $84.4 \pm 8.67$ & $87.8 \pm 7.87$ & $62.5 \pm 3.40$ & $69.8 \pm 5.12$ & $73.3 \pm 6.21$ \\
STANet \cite{Su2022}       & 0.04 & $80.8 \pm 6.43$          & $89.8 \pm 7.11$          & $91.7 \pm 5.89$          & $64.8 \pm 5.11$          & $72.1 \pm 6.98$          & $73.5 \pm 6.44$          \\
DenseNet-3D \cite{xu2024densenet}  & 0.76 & $-$             & $93.7 \pm 4.32$  & $95.3 \pm 4.44$  & $-$             & $-$             & $-$             \\
DBPNet \cite{ni2024dbpnet}       & 0.91 & $85.3 \pm 6.22$ & $94.4 \pm 4.62$ & $95.3 \pm 4.63$ & $74.0 \pm 5.20$ & $79.8 \pm 6.91$ & $80.2 \pm 6.79$ \\
DARNet \cite{Yan2024DARNet}       & 0.08 & $89.2 \pm 5.50$ & $94.8 \pm 4.53$ & $95.5 \pm 4.89$ & $74.6 \pm 6.09$ & $80.1 \pm 6.85$ & $81.2 \pm 6.34$ \\
MHANet \cite{Li2025MHANet}       & \textbf{0.02} & $90.2 \pm 4.83$ & $94.3 \pm 4.29$ & $95.9 \pm 3.67$ & $75.5 \pm 5.68$ & $82.2 \pm 8.13$ & $83.0 \pm 7.14$ \\
\midrule
\textbf{AFA-Net (Vanilla Att.)} & 0.03 & $86.9 \pm 5.62$ & $89.5 \pm 4.67$ & $92.1 \pm 4.1$ & $72.8 \pm 5.98$ & $75.3 \pm 6.32$ & $77.3 \pm 5.89$ \\
\textbf{AFA-Net (Diff. Att.)} & 0.03 & $\mathbf{92.1 \pm 5.34}^{*}$ & $\mathbf{95.2 \pm 4.25}^{*}$ & $\mathbf{96.8 \pm 3.59}^{*}$ & $\mathbf{78.1 \pm 6.03}^{*}$ & $\mathbf{82.6 \pm 5.47}$ & $\mathbf{85.1 \pm 5.68}^{*}$ \\
\bottomrule
\end{tabular}
\end{table*}

\begin{figure}[t]
  \centering
  \includegraphics[width=1.0\linewidth]{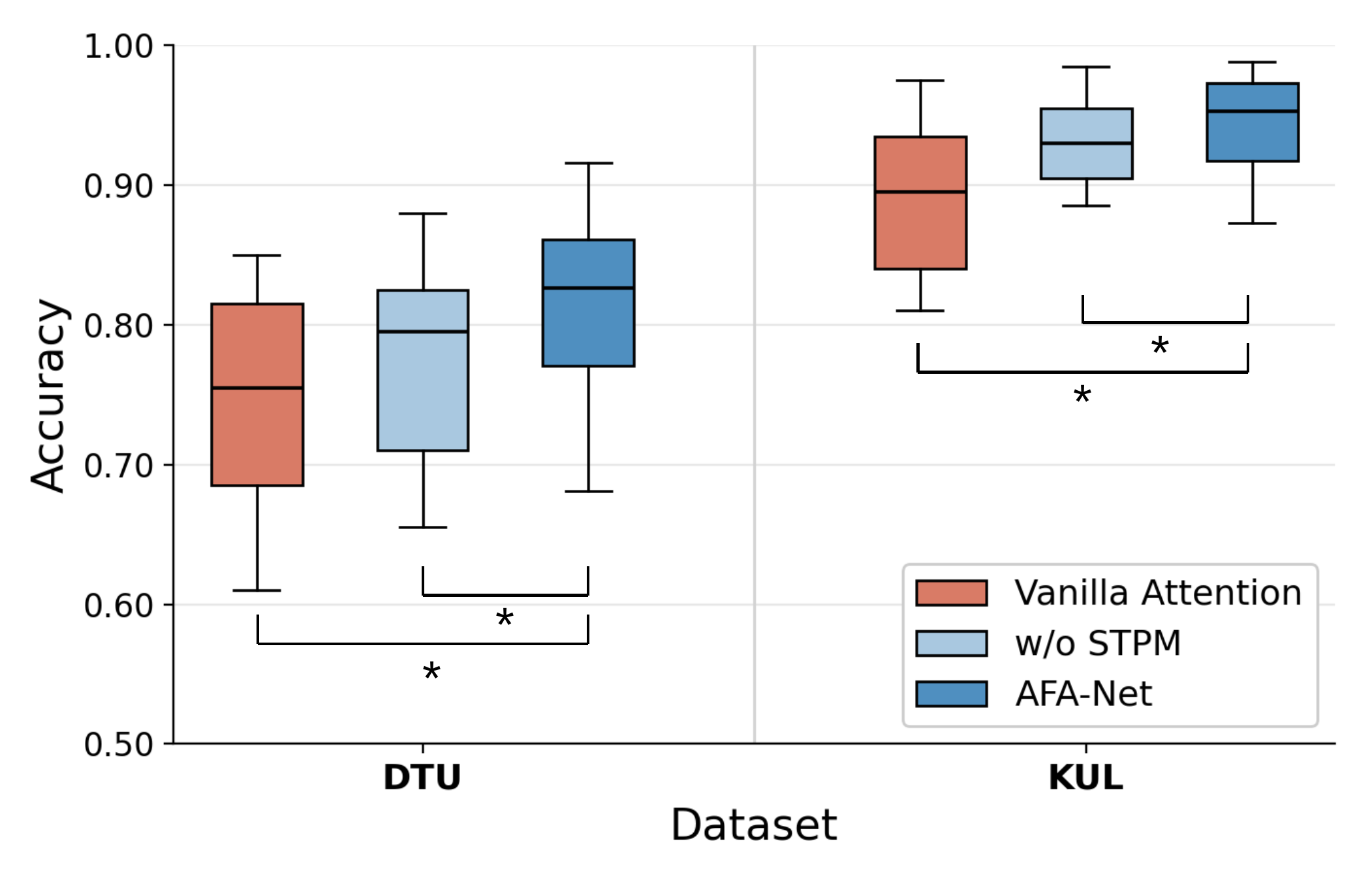}
  \caption{Ablation study to compare performance without differential attention and STPM for 1s decision window. The * symbol indicates a statistically significant difference for each component (two-tailed paired t-test, $p < 0.05$).}
  \label{fig:ablation}
\end{figure}

\vspace{-10pt}
\subsection{Ablation Study}
We evaluate all ablations using a 1s decision window, which most closely aligns with human attention switching \cite{jiang2022, fan2025}. All results are in Figure \ref{fig:ablation-bands} and Figure \ref{fig:ablation}.

Since cortical activity in the 1–10 Hz range carries the most relevant neural information in AAD \cite{ahissar2001speech, lalor2010neural}, we hypothesize that differential attention's gains come from exploiting these informative frequencies more effectively than vanilla attention. To test this, we remove individual frequency bands, \textit{delta} (1–4 Hz), \textit{theta} (4–8 Hz), \textit{alpha} (8–13 Hz), \textit{beta} (13–30 Hz), and \textit{gamma} (30–50 Hz) from only the test set, yielding five test variations, each compared to the test data with \textit{none} of the frequencies removed. As shown in Figure \ref{fig:ablation-bands}, differential attention's accuracy decreases more sharply when \textit{delta} and \textit{theta} are removed (roughly around 20–21\% on KUL and 13–19\% on DTU). Vanilla attention instead drops in accuracy more sporadically, with similarly large drops spread across the less informative \textit{alpha}, \textit{beta}, and \textit{gamma} bands (roughly around 17–22\% on KUL and 7–21\% on DTU). This confirms differential attention's gains stem more from prioritizing the frequency bands that carry the most information.

We also conduct an ablation study on AFA-Net's key components: replacing differential attention with vanilla attention and removing STPM. As shown in Figure \ref{fig:ablation}, both variants significantly underperform the full model on both datasets. Replacing differential attention with vanilla attention causes the largest drop, with the average accuracy falling by roughly 12.2\% on KUL and 7\% on DTU. Overall, both components contribute to AFA-Net's performance, with differential attention having the greater impact.

\vspace{-12pt}
\section{Conclusion}
In this paper, we proposed AFA-Net, a lightweight AAD architecture that utilizes differential attention to capture more relevant neural information. AFA-Net achieves competitive accuracy across all decision windows, with consistent performance gains over vanilla attention on both datasets. Overall, these results show that differential attention offers a simple, yet effective low-complexity path to more accurate AAD.

\footnotesize
\bibliographystyle{IEEEbib}
\bibliography{refs}

@article{wen2025hybrid,
  title={Hybrid channel attention network for auditory attention detection},
  author={Wen, Yahao and Ma, Shuai and Liu, Chuang and Wang, Yongjie},
  journal={Scientific Reports},
  volume={15},
  number={1},
  pages={38418},
  year={2025},
  publisher={Nature Publishing Group UK London}
}

@inproceedings{ICLR2025_00b67df2,
 author = {Ye, Tianzhu and Dong, Li and Xia, Yuqing and Sun, Yutao and Zhu, Yi and Huang, Gao and Wei, Furu},
 booktitle = {International Conference on Learning Representations},
 editor = {Y. Yue and A. Garg and N. Peng and F. Sha and R. Yu},
 pages = {144--164},
 title = {Differential Transformer},
 url = {https://proceedings.iclr.cc/paper_files/paper/2025/file/00b67df24009747e8bbed4c2c6f9c825-Paper-Conference.pdf},
 volume = {2025},
 year = {2025}
}

@article{vandecappelle2021eeg,
  title={EEG-based detection of the locus of auditory attention with convolutional neural networks},
  author={Vandecappelle, Servaas and Deckers, Lucas and Das, Neetha and Ansari, Amir Hossein and Bertrand, Alexander and Francart, Tom},
  journal={Elife},
  volume={10},
  pages={e56481},
  year={2021},
  publisher={eLife Sciences Publications, Ltd}
}

@article{OSullivan2015,
  author    = {James A. O'Sullivan and Alan J. Power and Nima Mesgarani and Siddharth Rajaram and John J. Foxe and Barbara G. Shinn-Cunningham and Malcolm Slaney and Shihab A. Shamma and Edmund C. Lalor},
  title     = {Attentional Selection in a Cocktail Party Environment Can Be Decoded from Single-Trial {EEG}},
  journal   = {Cerebral Cortex},
  volume    = {25},
  number    = {7},
  pages     = {1697--1706},
  year      = {2015},
  month     = {Jul},
  doi       = {10.1093/cercor/bht355},
  url       = {https://academic.oup.com/cercor/article/25/7/1697/457492}
}

@article{Mirkovic2015,
  author    = {Bojana Mirkovic and Stefan Debener and Manuela Jaeger and Maarten De Vos},
  title     = {Decoding the attended speech stream with multi-channel {EEG}: implications for online, daily-life applications},
  journal   = {Journal of Neural Engineering},
  volume    = {12},
  number    = {4},
  pages     = {046007},
  year      = {2015},
  doi       = {10.1088/1741-2560/12/4/046007},
  url       = {https://iopscience.iop.org/article/10.1088/1741-2560/12/4/046007}
}

@article{Mesgarani2012,
  author    = {Nima Mesgarani and Edward F. Chang},
  title     = {Selective cortical representation of attended speaker in multi-talker speech perception},
  journal   = {Nature},
  volume    = {485},
  number    = {7397},
  pages     = {233--236},
  year      = {2012},
  month     = {May},
  doi       = {10.1038/nature11020},
  url       = {https://www.nature.com/articles/nature11020}
}

@article{Golumbic2013,
  author    = {Elana M. Zion Golumbic and Nai Ding and Stephan Bickel and Peter Lakatos and Catherine A. Schevon and Guy M. McKhann and Robert R. Goodman and Ronald Emerson and Ashesh D. Mehta and Jonathan Z. Simon and David Poeppel and Charles E. Schroeder},
  title     = {Mechanisms underlying selective neuronal tracking of attended speech at a "cocktail party"},
  journal   = {Neuron},
  volume    = {77},
  number    = {5},
  pages     = {980--991},
  year      = {2013},
  doi       = {10.1016/j.neuron.2012.12.037},
  url       = {https://www.cell.com/neuron/fulltext/S0896-6273(12)01166-3}
}

@article{Ding2012,
  author    = {Nai Ding and Jonathan Z. Simon},
  title     = {Emergence of neural encoding of auditory objects while listening to competing speakers},
  journal   = {Proceedings of the National Academy of Sciences},
  volume    = {109},
  number    = {29},
  pages     = {11854--11859},
  year      = {2012},
  doi       = {10.1073/pnas.1205381109},
  url       = {https://www.pnas.org/doi/10.1073/pnas.1205381109}
}

@article{Akram2016,
  author    = {Sahar Akram and Alessandro Presacco and Jonathan Z. Simon and Shihab A. Shamma and Behtash Babadi},
  title     = {Robust decoding of selective auditory attention from {MEG} in a competing-speaker environment via state-space modeling},
  journal   = {NeuroImage},
  volume    = {124},
  pages     = {906--917},
  year      = {2016},
  doi       = {10.1016/j.neuroimage.2015.09.048},
  url       = {https://www.sciencedirect.com/science/article/abs/pii/S1053811915008708}
}

@article{Su2022,
  author    = {Enze Su and Siqi Cai and Longhan Xie and Haizhou Li and Tanja Schultz},
  title     = {{STAnet}: A Spatiotemporal Attention Network for Decoding Auditory Spatial Attention from {EEG}},
  journal   = {IEEE Transactions on Biomedical Engineering},
  volume    = {69},
  number    = {7},
  pages     = {2233--2242},
  year      = {2022},
  doi       = {10.1109/TBME.2021.3139645},
  url       = {https://ieeexplore.ieee.org/document/9664353}
}

@inproceedings{Kuruvila2021,
  author    = {Ivine Kuruvila and Jan Muncke and Eghart Fischer and Ulrich Hoppe},
  title     = {Extracting the auditory attention in a dual-speaker scenario from {EEG} using a joint {CNN-LSTM} model},
  booktitle = {Frontiers in Physiology},
  volume    = {12},
  year      = {2021},
  doi       = {10.3389/fphys.2021.700655},
  url       = {https://www.frontiersin.org/articles/10.3389/fphys.2021.700655/full}
}

@inproceedings{Li2025MHANet,
  author    = {Lu Li and Cunhang Fan and Hongyu Zhang and Jingjing Zhang and Xiaoke Yang and Jian Zhou and Zhao Lv},
  title     = {{MHANet}: Multi-scale Hybrid Attention Network for Auditory Attention Detection},
  booktitle = {Proceedings of the Thirty-Fourth International Joint Conference on Artificial Intelligence (IJCAI-25)},
  year      = {2025},
  url       = {https://www.ijcai.org/proceedings/2025/0465.pdf},
  note      = {arXiv:2505.15364}
}

@inproceedings{Yan2024DARNet,
  author    = {Shikha Pahuja and Debasis Samanta and Vipul Arora},
  title     = {{DARNet}: Dual Attention Refinement Network with Spatiotemporal Construction for Auditory Attention Detection},
  booktitle = {Proceedings of the 38th Conference on Neural Information Processing Systems (NeurIPS)},
  year      = {2024},
  url       = {https://neurips.cc/virtual/2024/poster/94483}
}

@article{Zhu2023,
  author    = {Jiawei Zhu and Enze Su and Siqi Cai and Chengwei Tong and Haizhou Li},
  title     = {{SSF-DST}: A Spectro-Spatial Features Enhanced Deep Spatiotemporal Network for {EEG}-Based Auditory Attention Detection},
  journal   = {IEEE Journal of Biomedical and Health Informatics},
  volume    = {27},
  number    = {12},
  pages     = {5786--5797},
  year      = {2023},
  doi       = {10.1109/JBHI.2023.3320208},
  url       = {https://ieeexplore.ieee.org/document/10262486}
}

@article{ramoser2000optimal,
  title={Optimal spatial filtering of single trial EEG during imagined hand movement},
  author={Ramoser, Herbert and Muller-Gerking, Johannes and Pfurtscheller, Gert},
  journal={IEEE transactions on rehabilitation engineering},
  volume={8},
  number={4},
  pages={441--446},
  year={2000},
  publisher={IEEE}
}

@article{blankertz2007optimizing,
  title={Optimizing spatial filters for robust EEG single-trial analysis},
  author={Blankertz, Benjamin and Tomioka, Ryota and Lemm, Steven and Kawanabe, Motoaki and Muller, Klaus-Robert},
  journal={IEEE Signal processing magazine},
  volume={25},
  number={1},
  pages={41--56},
  year={2007},
  publisher={IEEE}
}

@inproceedings{cai2021low,
  title={Low-latency auditory spatial attention detection based on spectro-spatial features from EEG},
  author={Cai, Siqi and Sun, Pengcheng and Schultz, Tanja and Li, Haizhou},
  booktitle={2021 43rd Annual International Conference of the IEEE Engineering in Medicine \& Biology Society (EMBC)},
  pages={5812--5815},
  year={2021},
  organization={IEEE}
}

@inproceedings{xu2024densenet,
  title={A DenseNet-based method for decoding auditory spatial attention with EEG},
  author={Xu, Xiran and Wang, Bo and Yan, Yujie and Wu, Xihong and Chen, Jing},
  booktitle={ICASSP 2024-2024 IEEE International Conference on Acoustics, Speech and Signal Processing (ICASSP)},
  pages={1946--1950},
  year={2024},
  organization={IEEE}
}

@inproceedings{ni2024dbpnet,
  title={DBPNet: Dual-Branch Parallel Network with Temporal-Frequency Fusion for Auditory Attention Detection.},
  author={Ni, Qinke and Zhang, Hongyu and Fan, Cunhang and Pei, Shengbing and Zhou, Chang and Lv, Zhao},
  booktitle={IJCAI},
  pages={3115--3123},
  year={2024}
}

@article{das2019auditory,
  title={Auditory attention detection dataset KULeuven},
  author={Das, Neetha and Francart, Tom and Bertrand, Alexander},
  journal={Zenodo},
  year={2019}
}

@article{fuglsang2018eeg,
  title={EEG and audio dataset for auditory attention decoding},
  author={Fuglsang, S{\o}ren A and Wong, DD and Hjortkj{\ae}r, Jens},
  journal={Zenodo},
  year={2018}
}

@article{vaswani2017attention,
  title={Attention is all you need},
  author={Vaswani, Ashish and Shazeer, Noam and Parmar, Niki and Uszkoreit, Jakob and Jones, Llion and Gomez, Aidan N and Kaiser, {\L}ukasz and Polosukhin, Illia},
  journal={Advances in neural information processing systems},
  volume={30},
  year={2017}
}

@article{Li2024RL,
  author = {Li, D. and Xie, L. and Wang, Z. and Yang, H.},
  title = {Brain Emotion Perception Inspired EEG Emotion Recognition With Deep Reinforcement Learning},
  journal = {IEEE Transactions on Neural Networks and Learning Systems},
  volume = {35},
  number = {9},
  pages = {12979--12992},
  year = {2024},
  doi = {10.1109/TNNLS.2023.3241285}
}

@article{Li2024Attention,
  author = {Li, Chao and Wang, Feng and Zhao, Ziping and Wang, Haishuai and Schuller, Bj{\"o}rn W.},
  title = {Attention-Based Temporal Graph Representation Learning for EEG-Based Emotion Recognition},
  journal = {IEEE Journal of Biomedical and Health Informatics},
  volume = {28},
  number = {10},
  pages = {5755--5767},
  year = {2024},
  doi = {10.1109/JBHI.2024.3392341}
}

@article{jiang2022,
  title={Detecting the locus of auditory attention based on the spectro-spatial-temporal analysis of {EEG}},
  author={Jiang, Yifan and Chen, Ning and Jin, Jing},
  journal={Journal of Neural Engineering},
  volume={19},
  number={5},
  pages={056035},
  year={2022},
  publisher={IOP Publishing}
}

@article{fan2025,
  title={Seeing helps hearing: A multi-modal dataset and a {M}amba-based dual branch parallel network for auditory attention decoding},
  author={Fan, Cunhang and Zhang, Hongyu and Ni, Qinke and Zhang, Jingjing and Tao, Jianhua and Zhou, Jian and Yi, Jiangyan and Lv, Zhao and Wu, Xiaopei},
  journal={Information Fusion},
  pages={102946},
  year={2025},
  publisher={Elsevier}
}

@article{ahissar2001speech,
  title={Speech comprehension is correlated with temporal response patterns recorded from auditory cortex},
  author={Ahissar, Ehud and Nagarajan, Srikantan and Ahissar, Merav and Protopapas, Athanassios and Mahncke, Henry and Merzenich, Michael M.},
  journal={Proceedings of the National Academy of Sciences},
  volume={98},
  number={23},
  pages={13367--13372},
  year={2001},
  publisher={National Academy of Sciences}
}

@article{lalor2010neural,
  title={Neural responses to uninterrupted natural speech can be extracted with precise temporal resolution},
  author={Lalor, Edmund C. and Foxe, John J.},
  journal={European Journal of Neuroscience},
  volume={31},
  number={1},
  pages={189--193},
  year={2010},
  publisher={Wiley}
}

\end{document}